\documentclass[10pt, conference, compsocconf]{IEEEtran}
\usepackage{graphicx}
\usepackage{booktabs}
\usepackage{bigstrut,bigdelim,multirow}
\usepackage{amssymb}
\usepackage{amsmath}
\usepackage{algorithm}
\usepackage{algorithmicx}
\usepackage{algpseudocode}
\usepackage{array}
\usepackage{float}
\usepackage{balance}
\usepackage{subfigure}
\usepackage[T1]{fontenc}
\usepackage{url}

\begin{document}
%
\title{Scalable RDF Data Compression using X10}


\author{
Long Cheng$^{\ast\ddagger}$,  Avinash Malik$^{\dagger}$, Spyros Kotoulas$^{\ddagger}$, Tomas E Ward$^{\ast}$, Georgios Theodoropoulos$^{\S}$ \\
\\
       $^{\ast}$National University of Ireland, Maynooth, Ireland\\
       $~^{\dagger}$University of Auckland, New Zealand\\
      $^{\ddagger}$IBM Research, Ireland\\
      $~^{\S}$Durham University, UK\\
%
}

\maketitle

\begin{abstract}
The Semantic Web comprises enormous volumes of semi-structured data
  elements. For interoperability, these elements are represented by long strings. 
  Such representations are not
  efficient for the purposes of Semantic Web applications that perform
  computations over large volumes of information. A typical  method
  for alleviating the impact of this problem is through the use of
  compression methods that produce more compact representations of the
  data. The use of dictionary encoding for this purpose is particularly
  prevalent in Semantic Web database systems. However, centralized
  implementations present performance bottlenecks, giving rise to the need for scalable, 
  efficient distributed encoding schemes. 
  In this paper, we describe an encoding implementation based on
  the asynchronous partitioned global address space (APGAS) parallel
  programming model. We evaluate performance on a cluster of up to 384 cores 
  and datasets of up to 11 billion triples (1.9 TB). Compared
  to the state-of-art MapReduce algorithm, we demonstrate a speedup of
  $2.6 - 7.4\times$ and excellent scalability. These results illustrate the
  strong potential of the APGAS model for efficient implementation of dictionary
  encoding and contributes to the engineering of larger scale Semantic Web applications.
\end{abstract}


%
\IEEEpeerreviewmaketitle

\section{Introduction}
\label{sec:introduction}

The Semantic Web is becoming mainstream. As Linked Data is increasingly published from domains such as general knowledge (DBpedia
\cite{Auer2007}), bioinformatics (Uniprot \cite{Uniprot}), and GIS (linkedgeodata \cite{SLHA11}), the potential for new knowledge synthesis and discovery increases
immensely. Capitalizing on this potential requires semantic web
applications which are capable of integrating the information available
from this rapidly expanding web. The web engineering challenges are currently pushing computing boundaries at exascale and beyond.

This web is build on the W3C's Resource Description Framework (RDF)
\cite{RDF} - a schema-less, graph-based data format which describes the
Linked Data model in the form of subject-predicate-object (SPO)
expressions based on the statement of resources and their
relationships. These expressions are known as RDF triples. For an
instance, the simple statement from DBpedia ($<$dbpedia:IBM$>$,
$<$dbpedia-owl:foundation-Place$>$, $<$dbpedia:New-York$>$) conveys the
information that the corporation IBM was founded in New York. The
Semantic Web already contains billions of such statements and this
number is growing rapidly. As the terms in a RDF statement consist of
long string characters in the form of either URIs or literals, storing
and retrieving such information directly on an underlying database
namely a triple store will result in (1) unnecessarily high disk-space
consumption and (2) poor query performance (querying on strings is
computationally intensive).

Dictionary encoding has been shown to be an efficient way to ameliorate
these problems. Using dictionary encoding all the terms are replaced by
numerical ids through a mapping dictionary, and all the original triples
are finally converted to id triples before storing. The conventional encoding approach is that all the terms retrieve their ids through sequential access of a single dictionary an approach which is easy to implement but not suitable for compressing large data sets due to time considerations and memory requirements. Consequently, encoding triples in parallel based on
a distributed architecture with multiple dictionaries, becomes an
attractive choice for this problem. 

In this paper, we propose a scalable solution for compressing massive RDF data in parallel. We develop an algorithm and implement it using the partitioned global address space (APGAS) model programming language - X10 \cite{Charles2005}. We evaluate performance with up to 384 cores and with datasets comprising of up to 11 billion triples (1.9 TB). Compared to the state-of-the-art~\cite{Urbani2013}, our approach is faster (by a factor of 2.6 to 7.4), can deal with incremental updates in an efficient manner (outperforming the state-of-the-art by several orders of magnitude) and supports both disk and in-memory processing.

The rest of this paper is organized as follows: Section~\ref{sec:related-work} provides a review of related work. Section~\ref{sec:challenge} presents the challenges of distributed implementation of RDF compression. Section~\ref{sec:rdf-compression} introduces the proposed RDF compression algorithm. Section~\ref{sec:improvements} discusses optimizations and improvements for the algorithm. Section~\ref{sec:exper-sett} describes the experimental framework while Section~\ref{sec:evaluation} provides a quantitative evaluation of the algorithm. Section~\ref{sec:concl-future-work} concludes the paper and points to directions for future work.

\section{Related Work}
\label{sec:related-work}

Compression has been extensively studied in various database systems, and has been considered as an effective way to reduce the data footprint and improve the overall query processing performance~\cite{Chen2001}~\cite{Abadi2006}. In terms of efficient storage and retrieval of RDF data, various approaches described in~\cite{Fernandez2010} are geared toward efficient storage and transfer, as opposed to having direct access to the data for efficient processing. RDF data compression as used with the most popular triple stores, such as RDF-3X~\cite{Neumann2008rdf3x}, is performed on the basis of a single dictionary table. This method does not avail of potential speed-up by parallel implementations. Various distributed solutions used to manage RDF data have been proposed in the literature~\cite{yars2}~\cite{HuangAR11}. Nevertheless, their main focus is on data distribution after all the statements have been encoded. There exists only two efficient methods focused on parallel compression of RDF data. One is based on parallel hashing~\cite{Goodman2011} and the other uses the MapReduce model~\cite{Urbani2013}.

Goodman et al.~\cite{Goodman2011} adapt the linear probing method
on their Cray XMT machine, and realize the parallel encoding on a single
dictionary through parallel hashing, exploiting specialized primitives of the Cray XMT. Their evaluation has shown that
their method is highly efficient and the run-time is linear with
the number of used cores. This method requires that all data is kept in memory and is deeply reliant on
the shared memory architecture of the Cray XMT, making it unsuitable for commodity distributed
memory systems. They report an improvement of 2.4 to 3.3 compared to the MapReduce system on an in-memory configuration. By
comparison, on similar datasets, our approach outperforms the MapReduce system a factor of 2.6 to 7.4, both on-disk and in-memory.

Compared with~\cite{Goodman2011}, the MapReduce method proposed by
Urbani et al.~\cite{Urbani2013} is more general in that it can be
run on ordinary clusters and on-disk. There are three main elements to their system: 
(1) the popular terms are cached in memory by sampling the
data set, so that these popular terms assigned to each task could be 
encoded locally and consequently prevent eventual load
balancing problems, (2) a hash function is used to assign grouped terms to
reduce tasks, which then assign the term identifier, keeping the consistency of the encoding, and (3) the
MapReduce framework facilitates the parallel execution of the program. 
Although their evaluation on Hadoop has shown that their system is efficient and scales well, as we will show in Section~\ref{sec:evaluation}, our approach is both faster and more flexible, exploiting the finer-grain control of the APGAS model.

\section{Challenges}
\label{sec:challenge}

In our distributed architecture, RDF data is partitioned and then compressed using a dictionary on each computation node. However, under this model there exist three main challenges:

\begin{itemize}
\item Consistency - a term appearing on different compute nodes should have the same id. 

\item Performance - ensuring consistency based on naive methods can lead
to serious performance degradation.

\item Load balancing - the heavy skew of terms which characterizes real world linked data~\cite{Kotoulas2010} may lead to hotspots for the nodes
responsible for encoding these popular terms.
\end{itemize}

Both in space and time, the mapping of a term need always keep its uniqueness. For example, once the term \emph{``dbpedia:IBM"} is first encoded as id \emph{``101"} on node A, when encoding this string on another node B, we should also use the same value \emph{``101"}. Hash functions are potentially useful, but the length of the hash required to avoid collisions when processing billions to terms makes the space cost prohibitive.

We can ensure the consistency of the compression in the above example by copying the mapping \emph{[dbpedia:IBM, 101]} from node A to node B, but network communication cost and dealing with concurrency (e.g. locking on data structures) would lead to low performance.

Compared with the two issues above, load balancing presents a bigger challenge as the distribution of terms in the Semantic Web is highly skewed: there exist both popular (like predefined RDF and RDFS vocabulary) and unpopular
terms (like identifiers for entities that only appear for a limited number of times). For a distributed system, like ours,
any compression algorithm needs to be carefully engineered so that good
network communication and computational load-balance are achieved. If
terms are assigned using a simple hash distribution algorithm, the
continuous re-distribution of all the terms would undoubtedly lead to an
overloaded network. Furthermore, popular terms would lead to load-balancing issues.

For the sake of explanation, let us categorize terms into three groups: high-popularity terms that appear in a significant portion of the input triples, low-popularity terms that appear less than a handful of times and average-popularity terms (which is also the largest portion of RDF data). The state-of-the-art MapReduce compression algorithm~\cite{Urbani2013}
efficiently processes high-popularity terms. The very first job in the
algorithm is to sample and assign identifiers to popular terms, using an arbitrarily chosen threshold. These identifiers as distributed to all nodes in the system, and assigned to terms locally at each node. This dramatically improves load balancing and speeds up computation. For the rest of the terms, the data is repartitioned, and identifiers are assigned. For low-popularity terms, this also works well, as there are not many redundant data transfers. For low-popularity terms, we can either retrieve their mappings (possibly for multiple nodes), or we can send the data to the node where it is going to be encoded. In either case, the number of messages will be limited. For medium-popularity terms, the situation is different: Assume a term that appears 10000 times, and we have 100 compute nodes. If all nodes would need to retrieve the mapping from a single node, we would need 200 messages. If we repartition the terms, we would need at least 10000 messages. One can easily see the situation reversed for a term that appears 100 times (i.e. partitioning data might be more efficient that retrieving mappings). How can we reconcile efficient encoding of popular and non-popular terms?

\section{RDF Compression}
\label{sec:rdf-compression}

In this section, we first present an overview of the X10 programming language used for our implementation. Then, we describe the details of our RDF compression algorithm.

\subsection{An Overview of X10}
\label{sec:an-overview-about}

X10~\cite{Charles2005} is a multi-paradigm programming language developed by IBM. It supports the asynchronous partitioned global address space (APGAS)
model and is specifically designed to increase programmer productivity,
while being amenable to programming shared memory and distributed memory
supercomputers. It uses the concepts of \texttt{place} and
\texttt{activity} as the kernel notions to exploit parallelism in the
available hardware. A place is a logical abstraction of the underlying
heterogeneous processing element in the hardware such as cores in a
multi-core architecture, GPUs, or a whole physical machine. Activities
are light-weight threads that run on places. X10 schedules activities on
places to best utilize the available parallelism. The number of places
is constant through the life-time of an X10 program and is initialized
at program startup. Activities on the other hand can be forked at
program execution time. Forking an activity can be blocking, wherein
the parent returns after the forked activity completes execution, or
non-blocking, where in the parent returns instantaneously, after forking an activity. Furthermore, these activities can be forked locally or on a remote place.

X10 provides an important data structure called distributed arrays
(\texttt{DistArray}) for programming parallel algorithms. It is very similar as an \texttt{Array}, except that they distribute information among multiple \textit{places} and one or more elements in the \texttt{DistArray} can be mapped to a single place using
the concept of points~\cite{Charles2005}. Additionally, we used the following three crucial parallel programming constructs for
our compression implementation.
\begin{itemize}
\item \texttt{at(p) S}: this construct executes statement \texttt{S}
  at a specific place \texttt{p}. The current activity is blocked until
  \texttt{S} finishes executing on \texttt{p}.
\item \texttt{async S}: a child activity is forked by this
  construct. The current activity returns immediately (non-blocking)
  after forking \texttt{S}.
\item \texttt{finish S}: this construct is used to block the current
  activity and then waits for all activities forked by \texttt{S} to
  terminate.
\end{itemize}

\subsection{Main Algorithm}
\label{sec:impl-compr}

\begin{figure}[!t]
    \centering
    \includegraphics[width=3.2in]{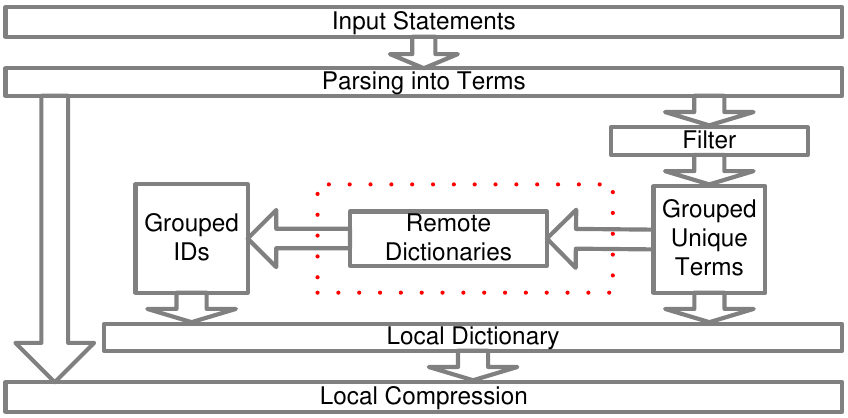}
    \caption{Data flow of the RDF Compression in our implementation.}
    \label{fig:dataflow}
\end{figure}

We describe the implementation of an RDF compression
algorithm on a distributed memory system.  We use distributed
dictionaries, one per place (recall that a place is a logical
abstraction for an underlying processing element), for encoding the
input data sets. Each data set is first divided into a number of
\textit{chunks} and assigned for processing on separate places. The initial partitioning of chunks is random. 
The overall implementation strategy for each place and the corresponding data flow 
are shown in Figure~\ref{fig:dataflow}.

First and foremost, every statement in the input set is parsed and split
into individual \textit{terms}, essentially, the \textit{subject}, the
\textit{predicate}, and the \textit{object}. All these parsed terms are filtered to remove the replications, and the extracted \textit{unique} terms are then divided into individual groups according to their hash values. The number of groups is set as the same as the number of places, and all terms in a aforementioned group have the same hash. In order to maintain consistency, the term' hash value maps it
to a \textit{single} dictionary in the distributed memory system where
it gets encoded. The groups of unique terms are pushed to the
dictionaries responsible for encoding these terms. Every place builds a
local dictionary, for encoding, based on the grouped unique terms and the corresponding group of ids received from remote nodes. Once all terms are encoded the grouped \textit{ids} are retrieved and the statements in the input data set are compressed.

\subsubsection*{Initialization}
\label{subsec:init}

We use the \texttt{DistArray} objects provided to implement our
distributed data structures. The initialization for these objects, at
each place, is shown in Algorithm~\ref{fig:encoding1}.

\begin{itemize}
\item \textit{dict} is the dictionary that maintains the term-id
  mappings during the whole compression process.
\item \textit{term\_c} collects the terms and keeps them in
  sequence for subsequent encoding.
\item \textit{local\_key\_c} is the array that collects the groups of
  unique terms that need to be sent to remote places for encoding.
\item \textit{local\_value\_c} is the array that collects all the
  encoded unique ids from remote places. The sequence of ids in
  \textit{local\_value\_c} is the same as terms in
  \textit{local\_key\_c}, thereby making it easy to insert the terms and
  their respective encodings into the local dictionary.
\item \textit{remote\_key\_c} is a temporary data structure used to
  receive the serialized the grouped unique terms that are sent from remote places.
\end{itemize}

\begin{algorithm}[!t]
\begin{algorithmic}[1]
    \State the number of places: $\emph{P}$
    \State Global initialize DistArray objects:
    \textit{dict} \textit{term\_c} \textit{local\_key\_c} \textit{local\_value\_c} \textit{remote\_key\_c}
    \State \texttt{finish} \texttt{async} \texttt{at} $p \in \mathit{P}
    $ \{ \\
    // \textit{here} the current place in X10
     \State $\textit{dict(here)}$:hashmap[string,long]
     \State $\textit{term\_c(here)}$:array[string] 
     \State $\textit{local\_key\_c(here)}$:array[array[string]]
     \State $\textit{local\_value\_c(here)}$:array[remote\_array[long]]
     \State $\textit{remote\_key\_c(here)}$:array[remote\_array[char]]\;
    \}
  \end{algorithmic}
  \caption{Initialization}
  \label{fig:encoding1}
\end{algorithm}

\subsubsection*{Term Grouping and Pushing}
\label{sec:push-terms}

We employ a \texttt{hashset} structure to process
the terms and to extract the unique terms that need to be transferred to
remote places. This is done for all terms irrespective of their
popularity. Using the \texttt{hashset} guarantees that any given term
can possibly move to a remote place just once, per current place.

The detailed implementation is given in Algorithm~\ref{fig:encoding2}. A
\texttt{hashset} is initialized at each place. Each \texttt{hashset}
collects terms according to their hash values. Before adding the parsed
term into the \textit{term\_c} queue, a term is added to the
\texttt{hashset}: \textit{key\_f}, if not already present. After
processing all the triples, the filtered terms will be copied into
\textit{local\_key\_c}, and then serialized and pushed to the assigned
place for further processing.

The structure \textit{local\_key\_c} is kept in memory for the later
local dictionary construction as shown in Figure~\ref{fig:dataflow}. The serialization/deserialization process is used only when the push array
objects are neither \texttt{long}, \texttt{int} nor \texttt{char}, otherwise we
directly transfer the data. Since the terms collected by each \texttt{hashset} are the unique ones to be sent
to remote places, the network communication and later computational
costs are significantly reduced. We use the \texttt{finish} operation in
this part to guarantee the completion of the data transfer at each place
before the term encoding.

\begin{algorithm}[!t]
  \begin{algorithmic}[1]
    \State \texttt{finish} \texttt{async} \texttt{at} $p \in
    \mathit{P}$ \{    
    \State Initialize $\textit{key\_f}$:array[hashset[string]]($P$) 
    \State Read in file $f_i$
    \For {$\textit{triple} \in f_i$}
    \State $\textit{terms}(3)$=parsing($\textit{triple}$)
    \For {$\textit{j} \gets 0..2$}
    \State $\textit{des}$=hash($\textit{terms}(j)$);
    \If {$\textit{terms}(j) \not \in \textit{key\_f}(des)$}  
    \State $\textit{key\_f}(des).add(\textit{term}(j))$
    \EndIf
    \State $\textit{term\_c}(here).add(\textit{term}(j))$
    \EndFor
    \EndFor
    \State Copy the terms in $\textit{key\_c}(i)$ to $\textit{local\_key\_c}(here)(i)$
    \For{$n \gets 0..(P-1)$} 
    \State Serialize $\textit{local\_key\_c}(here)(n)$ to $\textit{ser\_key}(n)$
    \State Push $\textit{ser\_key}(n)$ to $\textit{remote\_key\_c}(n)(here)$ at the place n
    \EndFor \ \}
  \end{algorithmic}
  \caption{Filter and Push Terms}
  \label{fig:encoding2}
\end{algorithm}

\subsubsection*{Term Encoding}
\label{sec:term-encoding}

Once the grouped unique terms have been transferred to the appropriate
remote places, the term encoding can commence. The term encoding
implementation at each place is similar to sequential encoding. The
received serialized char arrays, representing the grouped unique terms,
are deserialized to string arrays. Then the terms in such arrays access
the local dictionary sequentially to get their numerical ids. In this
process, if the mapping of a term already exists, its id is retrieved,
else, a new id is created, and the new mapping is added into the local
dictionary. In both cases, the id of the encoded term is added into a
temporary array for so that it can be sent back to the requester(s). The value of a new id is
determined by the summation of the largest id in the dictionary and the
value \textit{P}, the number of places. This guarantees there is no clash between term ids assigned at different places. Furthermore, each id is
formatted as an \texttt{unsigned} 64-bit integer in order to remove limitations regarding maximum dictionary size\footnote{it is possible to use arbitrary- or variable-length ids in order to further optimize space utilization, but this is beyond the scope of this paper.}.

We also write out the new mappings in this phase, as they build up the
final dictionary. Once the encoding of the grouped unique terms is
complete, we shift the \texttt{activity} to the corresponding place
where the terms originated, and retrieve the ids. We then proceed in processing
the following group. All encoding happens in parallel at each
place, and we use the \texttt{finish} operation synchronization. 
The details of the algorithm are given in Algorithm~\ref{fig:encoding3}.

\begin{algorithm}[!t]
  \begin{algorithmic}[1]
  \State \texttt{finish} \texttt{async} \texttt{at} $p \in P$ \{
   \State Initialize $\textit{key\_c}$:array[string], \textit{value\_c}:array[long]  
      \For {$i \gets 0..(P-1)$} 
       \State Deserialize $\textit{remote\_key\_c}(here)(i)$ to $\textit{key\_c}$
        \For {$\textit{key} \in \textit{key\_c}(i)$}
         \If {$\textit{key} \in \textit{dict}(here)$}
           \State $\textit{value\_c}.add(\textit{id})$
         \Else
           \State $\textit{id}=(\textit{dict}(here).\textit{size}+1)*P$  
           \State $\textit{dict}(here.id).\textit{put}(\textit{key,id})$
          \State $\textit{value\_c}.add(\textit{id})$
          \State Out-writing $<$\textit{key,id}$>$  
         \EndIf 
        \EndFor
        \State at $\textit{place}(i)$
       \State Pull $\textit{value\_c}(i)$ to $\textit{local\_value\_c}(here)(i)$
       \EndFor
       \ \}
  \end{algorithmic}
  \caption{Encode Terms and Pull Back IDs}
    \label{fig:encoding3}
\end{algorithm}

\subsubsection*{Statement Compression}
\label{sec:stat-compr}

The statements at each place can be compressed after all the ids of the
pushed terms have been pulled back. Since the terms and their respective
ids are held in order inside arrays, we can easily insert these mappings
into the local dictionary. Once inserted, we encode the parsed triples
in array \textit{term\_c}. Finally, we write out the ids to disk sequentially as shown in Algorithm~\ref{fig:encoding4}. The whole compression
process terminates when all individual activities terminate. Note that, in the actual implementation, we build a temporary hashmap to hold all the mappings and discard it after the encoding to optimize memory use.

\begin{algorithm}[!t]
  \begin{algorithmic}[1]
    \State \texttt{finish} \texttt{async} \texttt{at} $p \in P$ \{
    \For {$i \gets 0..(P-1)$} \State Add $<$\textit{key,id}$>$ from
    $\textit{local\_key\_c}(here)(i)$ and
    $\textit{local\_value\_c}(here)(i)$ to $\textit{dict}(here)$
    \EndFor
    \For{$\textit{term} \in \textit{term\_c}(here)$}
    \State $\textit{id} = \textit{dict}(here).get(\textit{term}).hashcode()$
    \State Out-writing $\textit{id}$
    \EndFor  \ \}
  \end{algorithmic}
  \caption{Statement Compression}
   \label{fig:encoding4}
\end{algorithm}

\section{Improvements}
\label{sec:improvements}

In this section, we present a set of extensions to our basic algorithm which improve efficiency and extend the applicability of the approach to a larger set of problems and computation platforms. The section concludes with a brief account of the theoretical complexity of our algorithm.

\subsection{I/O and Data Transfers}
\label{sec:io-data-transferring}

X10 does not yet provide efficient I/O
operation libraries for reading large data sets, as noted by Zhang et
al.~\cite{Zhang2011}. Moreover, using the standard \texttt{at\{p\}}
construct for copying data incurs a substantial penalty for deep copying data structures. In
order to alleviate these bottlenecks, Zhang et al., recommend the use of
\texttt{mmap} system call and \texttt{array.asycCopy} method. We adopt
the latter approach and extend the first one with the
\texttt{zlib} compression library to provide more efficient
reading of large data sets.

Our preliminary experiments suggest that using \textit{just} the
\texttt{mmap} approach for large I/O operations scales well to medium
sized data sets with less than hundreds of gigabytes of data. However,
for very large data sets measured in tera-bytes, reading \texttt{gzip}-compressed files in memory and decompressing them on the fly
results in substantially improved I/O performance. Moreover, compressing
data in the \texttt{gzip} format also reduces disk space usage.

The X10 standard library does not provide any interface for reading and
writing compressed gzip files, so we build a small library based on
\texttt{zlib} and integrate it with our X10 code via the foreign
function interface. We use the compressed datasets only while reading, since the resultant output is comparatively small and we simply write it out in bytes using the \texttt{OutputStreamWriter} class in
X10 standard library.

\subsection{Flexible Memory Footprint}
\label{subsec:flex-memory-footpr}

In our algorithm, the \texttt{DistArray} objects (Figure~\ref{fig:dataflow})
are kept in memory throughout the compression process. This limits the applicability of the method to clusters with sufficient memory to hold all data structures in memory.

To alleviate this problem, we divide the input data set into multiple \textit{chunks}, usually a multiple of the number of places. The
corresponding code change is shown in Algorithm~\ref{fig:loop}. The encoding process is divided into multiple loop iterations
corresponding to each chunk. In each of these compression iterations,
a place is assigned a specified number of chunks (line 2), while the
local \texttt{DistArray} objects are reused. This method makes our algorithm suitable for nodes with various memory sizes, provided the chunks are small enough. Note that the chunks can be made smaller by simply dividing the input data set into more chunks. It is expected that too many such chunks would lead to a decrease in performance, as there would be redundant filter and push operations for the same terms at the same place in different loops. We assess this trade-off through the evaluation in Section~\ref{sec:scalability}.

\begin{algorithm}[!t]
  \begin{algorithmic}[1]  
    \For {$i \gets 0..(loop-1)$}  
    \State Assign each place \textit{c} data chunks
    \State Parallel processing at each place
    \EndFor
  \end{algorithmic}
  \caption{Processing Data Chunks in Loops}
  \label{fig:loop}
\end{algorithm}

\subsection{Transactional Data Processing}
\label{sec:trans-data-proc}

A commonly occurring scenario is real-time processing of RDF data
sets. In such cases, data is inserted as part of a \textit{transaction}, and normally the chunks of data inserted are very small containing only a few hundred statements. In such a scenario,
there is no need to distribute data sets. Instead, one could just
compress the data set using a single cluster node. In our prototype, the
number of cluster nodes is controlled by the \texttt{X10\_NPLACES}
option. Furthermore, parallel transactions with multiple data sets on
multiple nodes are also supported using the same option. Finally, an
optimized data-node assignment strategy can be integrated with our
implementation if needed, but such a strategy is out of the scope
of this paper. Similarly, in this paper, we do not address rolling back transactions or deletes. 
In general, although our system can be extended to support transactional loads, its main utility is in encoding large datasets.

\subsection{Incremental Update}
\label{sec:incremental-update}

Another typical application is the incremental update of RDF data
sets. It is often required that such systems must encode a new dataset as an increment to already encoded datasets. Typically, the new input data set is
large. In this scenario, local dictionaries could be read in memory
before the encoding process. The extension of our algorithms for incremental update is shown in Algorithm~\ref{fig:update}.

\begin{algorithm}[!t]
  \begin{algorithmic}[1]  
    \State \texttt{finish} \texttt{async} \texttt{at} $p \in
    \mathit{P}$ \{
    \For {$\textit{$<$key,id$>$} \in \textit{local\_dict}$} 
    \State $\textit{table}(here.id).add(\textit{key,id})$
    \EndFor
    \State Processing new data\   
    \}  
  \end{algorithmic}
  \caption{Processing Update}
  \label{fig:update}
\end{algorithm}

\subsection{Algorithmic Complexity}
\label{sec:algor-compl}

Our compression algorithm with the aforementioned improvements has a
worst case computational complexity linear in the number of statements
of the input datasets $O(|N|)$ and the number of places $O(|P|)$. Herein, we describe the formulation of
our worst case complexity.

For a given place, the worst case complexity of the algorithm is $|P|$,
where $|P|$ is the number of places. This complexity is determined by
the largest loop at line 13 in Algorithm~\ref{fig:encoding2}. The total complexity of the algorithm is $O(|P|\times|P|\times|loop|/|P|)$, because there are a total of
$|P|$ places and all their implementations are nested inside the
\texttt{loop} variable in Figure~\ref{fig:loop}. The divisor
($|P|$) arises because each of these loops run in parallel. Therefore, the
overall worst case complexity is ($O(|loop|\times|P|)$). Based on this,
(a) for a constant number of places, the complexity of the algorithm is:
$O(|loop|)$, hence, the complexity of the algorithm is linear in the
value of loop. Next, if the size of each chunk is fixed,
assuming \texttt{k} triples per chunk and the total number of triples
are \texttt{N}, then the \texttt{loop} would be
($|\emph{N}|/|\emph{k}|/|\emph{P}|$). Thus, the complexity of the
algorithm will be $O(N)$, namely linear with the number of input triples
\texttt{N}, and (b) similarly, for a constant input size, the complexity
of the algorithm will be $O(P)$ linear in the number of places or cores
in the underlying execution architecture, provided each logical place is
mapped to a single core (as in our case).

\section{Experimental setup}
\label{sec:exper-sett}

We have conducted a rigorous quantitative evaluation of the proposed
encoding based on the setup as follows.

\subsection{Platform}
\label{sec:platform}

Our evaluation platform was the \emph{Exascale Systems Research Cluster}
in IBM Research Ireland. Each computation unit of this
cluster is an iDataPlex node with 2 Intel Xeon X5679 processors each
with 6 hardware cores running at 2.93 GHz, resulting in a total of 12
cores per physical node. Each node has 128GB of RAM and a single 1TB
SATA hard-drive. Nodes are connected by Gigabit Ethernet switch. The
operating system is Linux kernel version 2.6.32-220 and the software
stack consists of Java version 1.6.0\_25 and gcc version 4.4.6.

\subsection{Setup}
\label{sec:setup}

We have used X10 version 2.3 compiled to C++ code. We set
the \texttt{X10\_NPLACES} to the number of cores and the
\texttt{X10\_NTHREADS} to $1$, namely, one activity per place, which
avoids the overhead of context switching at runtime.

We compare our results with the MapReduce compression programme first described in~\cite{Urbani2013}. We use the latest version and run it on Hadoop
v0.20.2. We set the following system parameters: \emph{map.tasks.maximum} and
\emph{reduce.tasks.maximum} to 12, the \emph{mapred.child.java.opts} to
2 GB and the rest of the parameters are left to the default values.
The implementation parameters are configured with the
recommended values: \emph{samplingPercentage} is set to 10,
\emph{samplingThreshold} to 50000 and \emph{reducetasks} to the number
of cores. We have verified the suitability of these settings with the authors. 

We empty the file system cache between tests to minimize the effects of caching by the operating system are run the test three times, recording average values. 

\subsection{Datasets}
For our evaluation, we have used a set of real-world and benchmark datasets (as Table~\ref{tab:1}): DBpedia~\cite{Auer2007} is an extract of the
structured information from Wikipedia pages represented in RDF triples. LUBM~\cite{Guo2005} is a widely used benchmark
that can generate RDF data sets of arbitrary size. BTC~\cite{btc} is
a Web crawl encoding statements as N-Quads, while
Uniprot~\cite{Uniprot} is a large collection of biological function of
proteins derived from the research literature. We chose these data sets
because they vary widely in terms of size and kind of data they represent. The popularity and diversity of these datasets contributes to an unbiased evaluation.

\section{Evaluation}
\label{sec:evaluation}

We divide the presentation of our evaluation into different
sections. Section~\ref{sec:runtime}, compares the runtime and compression performance of our algorithm against the MapReduce implementation~\cite{Urbani2013}. We also evaluate the runtime
performance of our algorithm for the transactional and incremental update scenarios as described previously. Section~\ref{sec:scalability} examines the scalability of our algorithm and compares it against the scalability achieved by the MapReduce approach for increasing both numbers of processing units and input data set size. Finally, we present the load-balancing characteristics of our system in Section~\ref{sec:load-balancing}.

\subsection{Runtime}
\label{sec:runtime}

\subsubsection*{Compression} We perform the encoding using 16 nodes (192 cores) and report the compression results achieved by our algorithm in Table~\ref{tab:1}: Column \textit{\# Stats} gives the number of statements (triples) in each benchmark. The size of the input data sets is given both in the
terms of plain and gzip format in columns 3
and 4. The output column is composed of the compressed
statements and the corresponding dictionary tables at all places. Finally, the resulting compression ratio is calculated by dividing the size of the input files (in plain format) by the size of the total output. The compression ratios for the four data sets are similar: in the range of $4.1 - 4.5$. Note that although these ratios are smaller than the compression ratio achieved by \texttt{gzip}, our output data can be processed directly and we can also compress these outputs further using \texttt{gzip}, if need be. We achieve smaller compression ratios compared to MapReduce~\cite{Urbani2013}, because we use 64-bit integers to encode all terms, while their approach uses smaller integers for encoding parts of terms as well as further \texttt{gzip} compression on their output data.

\begin{table}[!t]
  \centering
  \caption{Dataset information and compression achieved}
  \begin{tabular*}{\columnwidth}{@{\extracolsep{\fill}}ccccccc}
    \toprule[0.8pt]
    \multirow{2}[4]{*}{\textbf{Dataset}} & \multirow{2}[4]{*}{\textbf{\# Stats.}} & \multicolumn{2}{c}{\textbf{Input (GB)}} & \multicolumn{2}{c}{\textbf{Output (GB)}} & \textbf{Compr.} \\
    &       & \textbf{Plain} & \textbf{Gzip} & \textbf{Data} & \textbf{Dict.} & \textbf{Ratio} \\
    \midrule
    DBpedia & 153M  & 25.1  & 3.5   & 3.5   & 2.7   & 4.1 \\
    LUBM  & 1.1B  & 190   & 5.5   & 24.8  & 17.7  & 4.5 \\
    BTC2011 & 2.2B  & 450   & 20.9  & 65.6  & 40    & 4.3 \\
    Uniprot & 6.1B  & 797   & 58.7  & 136   & 46.4  & 4.4 \\
    \bottomrule[0.8pt]
  \end{tabular*}%
  \label{tab:1}
\end{table}%

\subsubsection*{Runtime and Throughput} We compare the runtime and throughput between our approach and that of
the MapReduce framework in two cases: disk-based and
in-memory compression. In the first case, the reading and writing data
is on disk (or HDFS based on disk). For the latter, we
process all data in memory. For memory based I/O, we pre-read the
statements in an \texttt{ArrayList} at each place and also assign the
output to \texttt{ArrayList}. As MapReduce does not provide such
mechanisms, we instead set the path of the Hadoop parameter
\textit{hadoop.tmp.dir} to a \texttt{tmpfs} file system resident in memory. The results of
these two cases are shown in Table~\ref{tab:2a} and Table~\ref{tab:2b}. We define runtime as the time taken for the whole encoding process: reading files, performing encoding and writing out the
compressed triples and dictionaries. The throughput is described in terms of two aspects: (a) rate, which is calculated by dividing the input size (in plain format) by the algorithm runtime, and (b) statements processed per second that is calculated by dividing the number of processed statements by the runtime.

\begin{figure}[!t]
  \centering
  \includegraphics[width=\columnwidth]{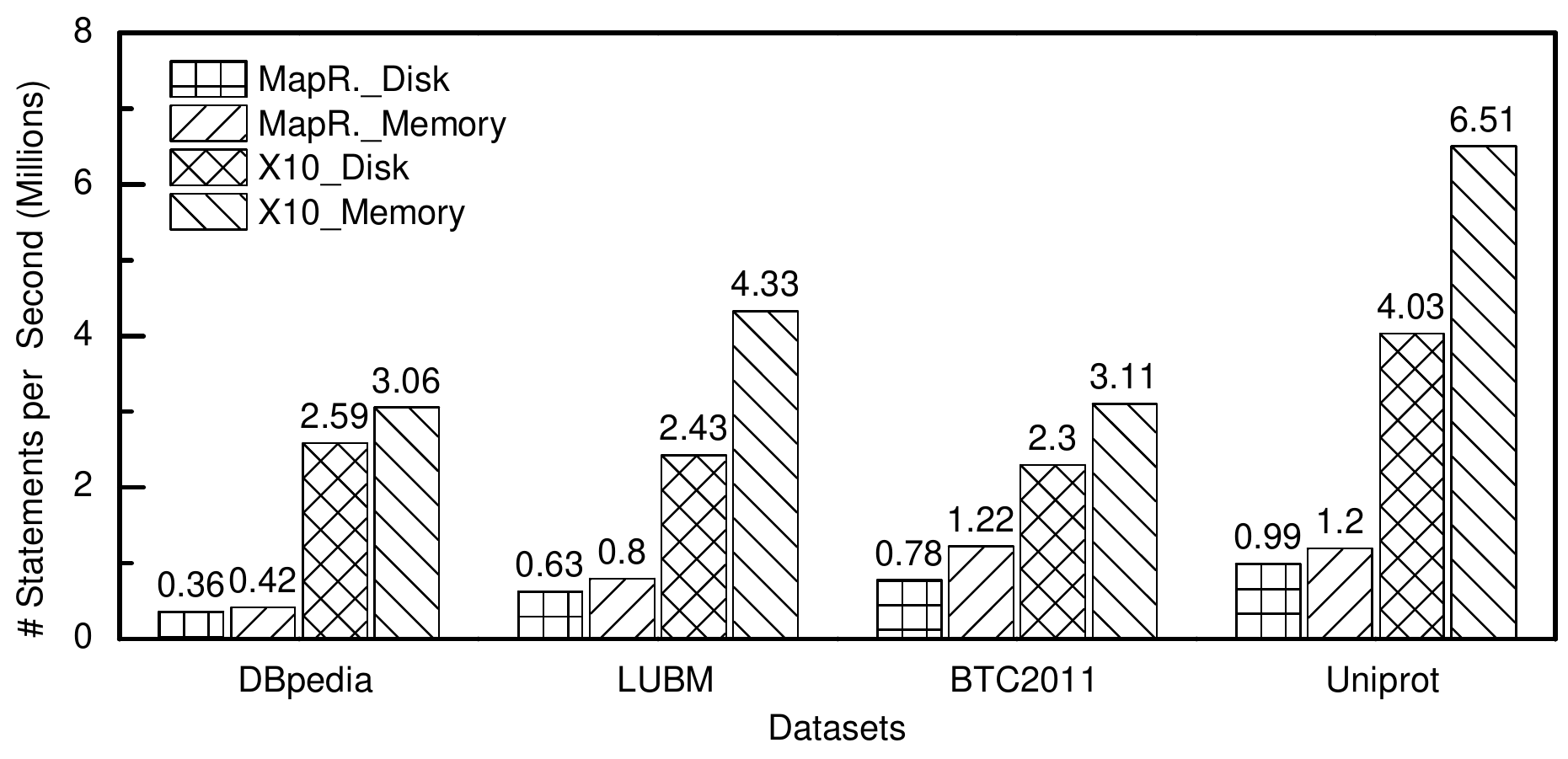}
  \caption{Throughput of the two implementations using 192 cores, based on disk-based and memory-based cases with the four datasets.}
  \label{fig:throughput}
\end{figure}

From Table~\ref{tab:2a}, our approach is $2.9-7.3\times$
faster than the MapReduce-based approach for disk-based computation, and $2.6-7.4\times$ for in-memory as illustrated in
Table~\ref{tab:2b}. The smallest speedup occurs for the BTC2011 benchmark, however it should be noted that in this instance, whereas we compress \texttt{N-Quads}, MapReduce discards the fourth term in the input data and just compresses the first three terms. Moreover, the compression throughput of Uniprot in both cases is much higher than the other three datasets. We attribute this to the large number
of recurring popular terms. Comparing the two cases, the in-memory compression is faster than the disk-based one for both
algorithms, although not dramatically so. Moreover, the improvements we
achieved in Table~\ref{tab:2b} are greater than those in
Table~\ref{tab:2a} for the LUBM and Uniprot data sets, marginally greater
for DBpedia and slightly smaller for the BTC2011 data set. This illustrates that the two algorithms gain disproportionally from the faster I/O over different data sets (with our system showing better gains overall). Figure~\ref{fig:throughput} shows that the maximum number of statements processed per second is about 6.51M, higher than any method in the literature. 

\begin{table}[!t]
  \centering
  \caption{Disk-based Runtime and Rates of Compression (192 cores)}
  \begin{tabular*}{\columnwidth}{@{\extracolsep{\fill}}cccccc}
    \toprule[0.8pt]
    \multirow{2}[4]{*}{\textbf{Dataset}} & \multicolumn{2}{c}{\textbf{Runtime (sec.)}} & \multicolumn{2}{c}{\textbf{Rates (MB/s)}} & \multirow{2}[4]{*}{\textbf{Imprv.}} \\
    & \textbf{MapR.} & \textbf{X10} & \textbf{MapR.} & \textbf{X10} &  \\
    \midrule
    DBpedia & 430   & 59    & 59.7  & 435   & 7.3 \\
    LUBM  & 1739  & 453   & 111.9 & 429.5 & 3.8 \\
    BTC2011 & 2817  & 956   & 163.6 & 482   & 2.9 \\
    Uniprot & 6160  & 1515  & 132.5 & 538.7 & 4.0 \\
    \bottomrule[0.8pt]
  \end{tabular*}%
  \label{tab:2a}%
\end{table}%

\begin{table}[!t]
  \centering
   \caption{In-memory Runtime and Rates of Compression (192 cores)}
    \begin{tabular*}{\columnwidth}{@{\extracolsep{\fill}}cccccc}
    \toprule[0.8pt]
    \multirow{2}[4]{*}{\textbf{Dataset}} & \multicolumn{2}{c}{\textbf{Runtime (sec.)}} & \multicolumn{2}{c}{\textbf{Rates (MB/s)}} & \multirow{2}[4]{*}{\textbf{Imprv.}} \\
          & \textbf{MapR.} & \textbf{X10} & \textbf{MapR.} & \textbf{X10} &  \\
              \midrule
    DBpedia & 368   & 50    & 69.8  & 514 & 7.4 \\
    LUBM  & 1382  & 254   & 140.8 & 766 & 5.4 \\
    BTC2011 & 1809  & 708   & 254.7 & 650.8 & 2.6 \\
    Uniprot & 5076  & 937   & 160.8 & 871   & 5.4 \\
    \bottomrule[0.8pt]
    \end{tabular*}
  \label{tab:2b}%
\end{table}%

\subsubsection*{Transactional} We simulated two transactional processing scenarios with in-memory compression: (1) sequential
transactions on a single node and (2) multiple parallel transactions on
multiple nodes using the LUBM data set. To
simulate transactions, we first encode the 1.1 billion triples in
the \textit{LUBM8000} benchmark. Next, we prepare a RDF data set that
contains 1M triples, split into 10K, 1K, 100, and 10 chunks,
respectively. After encoding is complete, we  encode these new input chunks (every 10 chunks) sequentially and
record the corresponding encoding time. For the multiple parallel
transaction scenario, we could only record the encoding time for our
implementation since Hadoop uses a centralized model for data storage.

Results are presented in Table~\ref{tab:3}. One can clearly observe
that our approach is orders of magnitude faster than the MapReduce
approach for the sequential case. The latter is neither optimized nor suitable for this use-case, since the startup overhead dominates the runtime, as evident from the observation that the average time to process chunks with different sizes is approximately the same.
For our system, we observe that the average runtime of our approach increases with increasing chunk sizes, and the trend moves toward linear for the sequential case. This means that, for a single place, overhead takes a larger proportion of the runtime.

Since we are using 192 cores and the number of chunks used in this scenario is
10, for each transaction with the parallel processing by our prototype, the chunks can be compressed at once by 10 places in parallel. The results
in Table~\ref{tab:3} show that the runtime is around 0.2 seconds when the number of statements is less than 100 in each chunk, which
is slightly worse than our expectations for real-time applications, although still well within an acceptable range. Upon further analysis, we have found that
this increase in program runtime is due to underlying bottlenecks
in the X10 runtime implementation, which we have not addressed in this paper: (a)
Every \texttt{async} call forks an underlying \texttt{pthread} (Posix
thread) \textit{atomically}, which leads to execution time overhead. (b)
Type initializations in X10 are expensive, because all type
initializations are internally guarded by locks. Our implementation
still performs reasonably well even with these implementation overheads.

\begin{table}[!t]
  \centering
  \caption{Processing 1M Statements in the Transactional Scenario}
  \begin{tabular*}{\columnwidth}{@{\extracolsep{\fill}}cccc}
    \toprule[0.8pt]
    \textbf{\# Stats} & \multicolumn{3}{c}{\textbf{\textit{Avg. runtime per 10 chunks (sec.)}}} \\
    \textbf{per chunk} & \textbf{MapR.} & \textbf{X10} & \textbf{X10\_Para.} \\
        \midrule
    100   & 439   & 0.211 & 0.164 \\
    1K    & 441   & 0.359 & 0.391 \\
    10K   & 454   & 1.761 & 0.648 \\
    100K  & 454   & 17.177 & 2.192 \\
    \bottomrule[0.8pt]
  \end{tabular*}%
  \label{tab:3}%
\end{table}%

\subsubsection*{Updates} We evaluate the incremental updates scenario for RDF
compression again using the \textit{LUBM8000} dataset and by splitting
it into 2, 4, and 8 chunks, respectively. The resulting datasets are
compressed in 2, 4 and 8 different executions respectively. Before each compression cycle, we empty the cache as to simulate real world conditions. The results
comparing our approach and MapReduce are shown in Table~\ref{tab:4}. As
expected, the performance for both algorithms decreases with increasing
number of chunks, because of the additional process required during the
encoding (e.g. reading the dictionary into memory). However, the
increase in program runtime for our approach is much smaller than
MapReduce. A possible explanation is that because our dictionary reading operation is faster, the startup overhead of our system is lower. It is also possible that the efficacy of the popularity caching technique used by MapReduce decreases disproportionately as the number of chunks increases. 


\begin{table}[!t]
  \centering
  \caption{Incremental Update Scenario with different chunk size}
  \begin{tabular*}{\columnwidth}{@{\extracolsep{\fill}}ccccc}
    \toprule[0.8pt]
    \multirow{2}[4]{*}{\textbf{\#  Chunks}} & \multirow{2}[4]{*}{\textbf{Chunk Size}} & \multicolumn{2}{c}{\textbf{Runtime (sec.)}} & \multirow{2}[4]{*}{\textbf{Imprv.}} \\
    &       & \textbf{MapR.} & \textbf{X10} &  \\
    \midrule
    1     & 190 GB & 1739  & 453   & 3.8 \\
    2     & 95 GB & 2468  & 551   & 4.5 \\
    4     & 47 GB  & 3900  & 755   & 5.2 \\
    8     & 23 GB & 6704  & 1164  & 5.8 \\
    \bottomrule[0.8pt]
  \end{tabular*}%
  \label{tab:4}%
\end{table}%

\subsection{Scalability}
\label{sec:scalability}

\begin{figure*}[!t]
\center
\subfigure[Runtime by varying nodes]{ 
\label{fig:subfig:1a}
\includegraphics[width=2.21in]{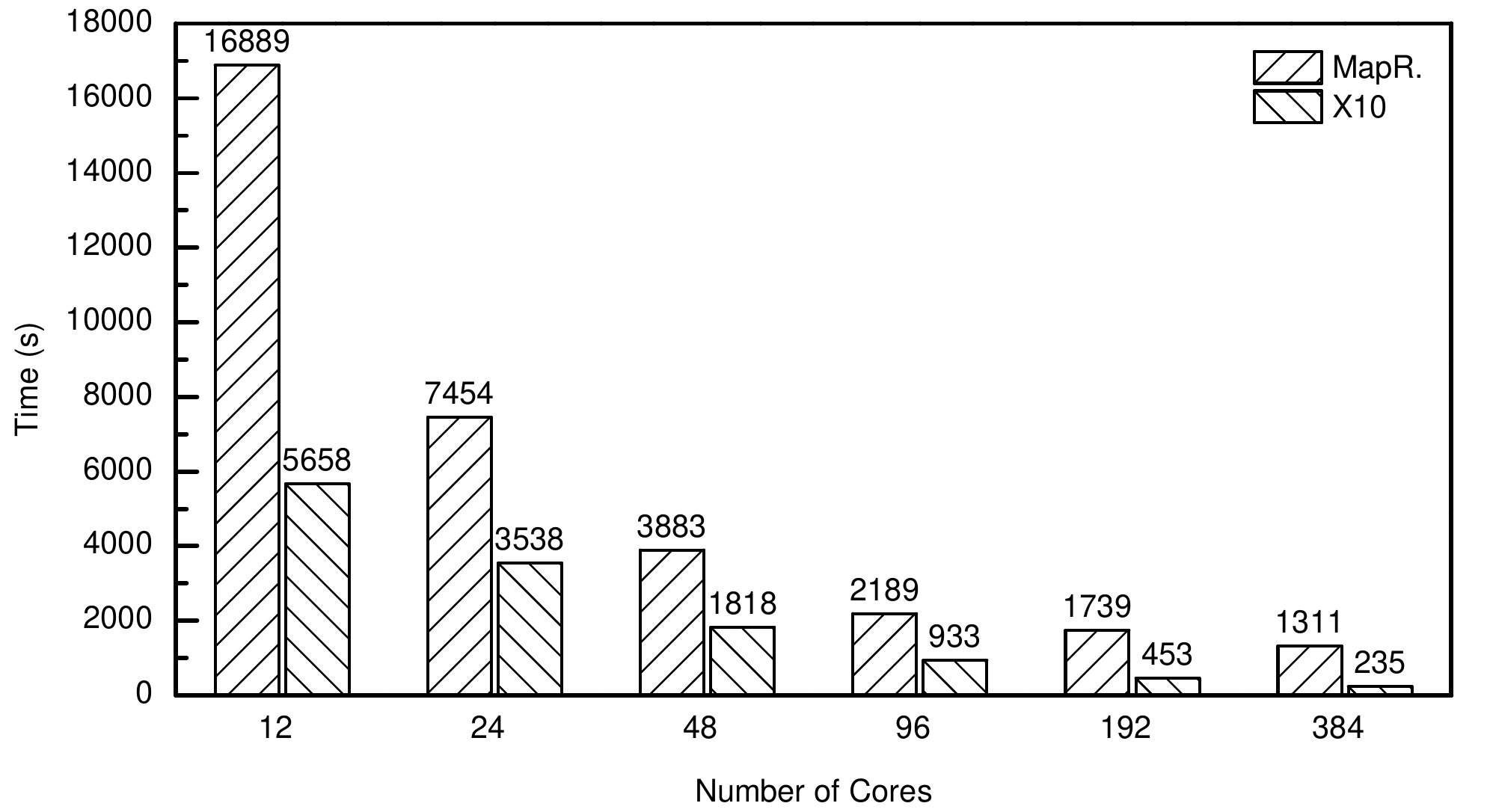} }
\subfigure[Speedups by varying nodes]{ 
\label{fig:subfig:1b}
\includegraphics[width=2.2in]{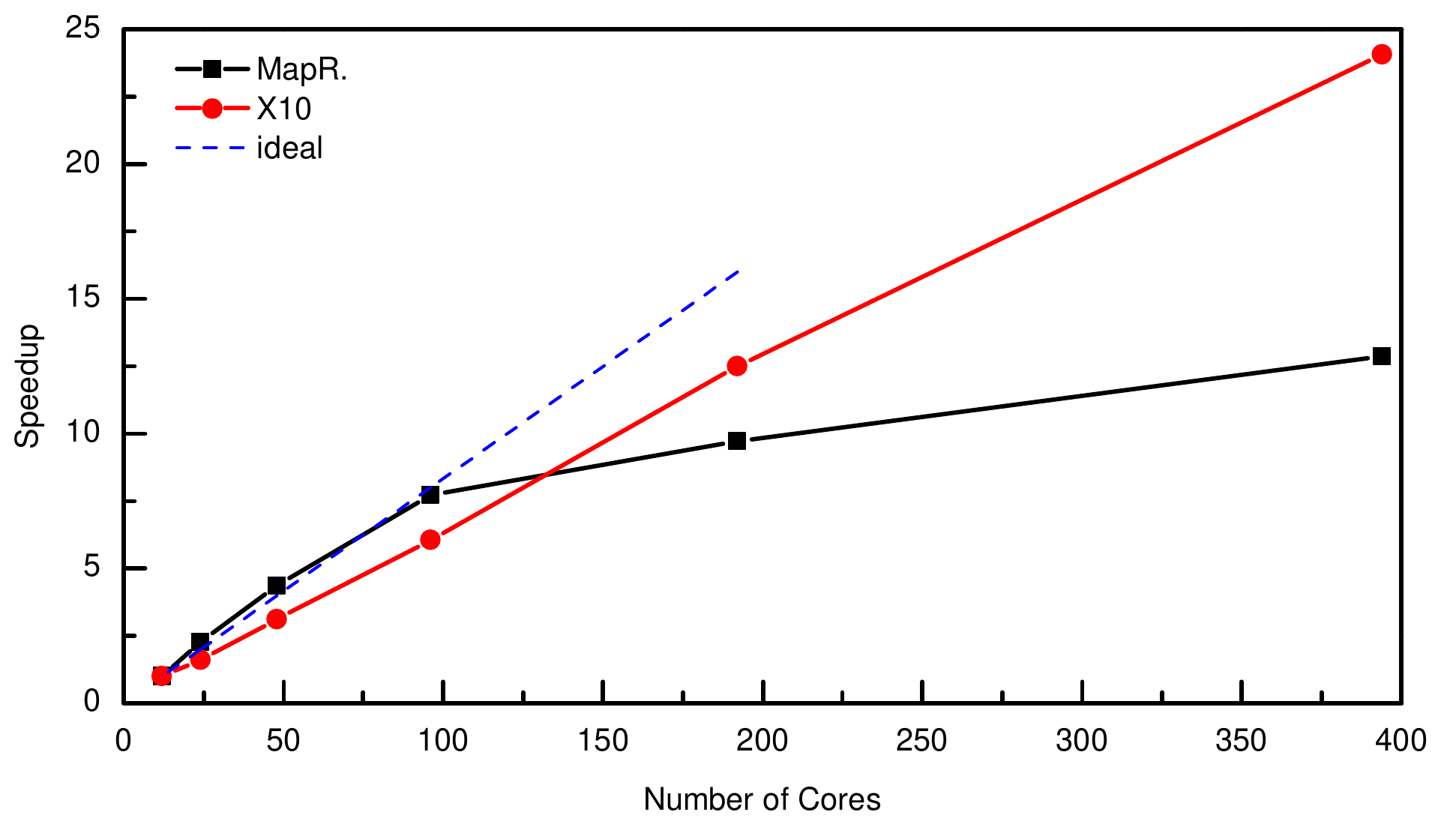} }
\subfigure[Runtime by varying size]{ 
\label{fig:subfig:1c}
\includegraphics[width=2.2in]{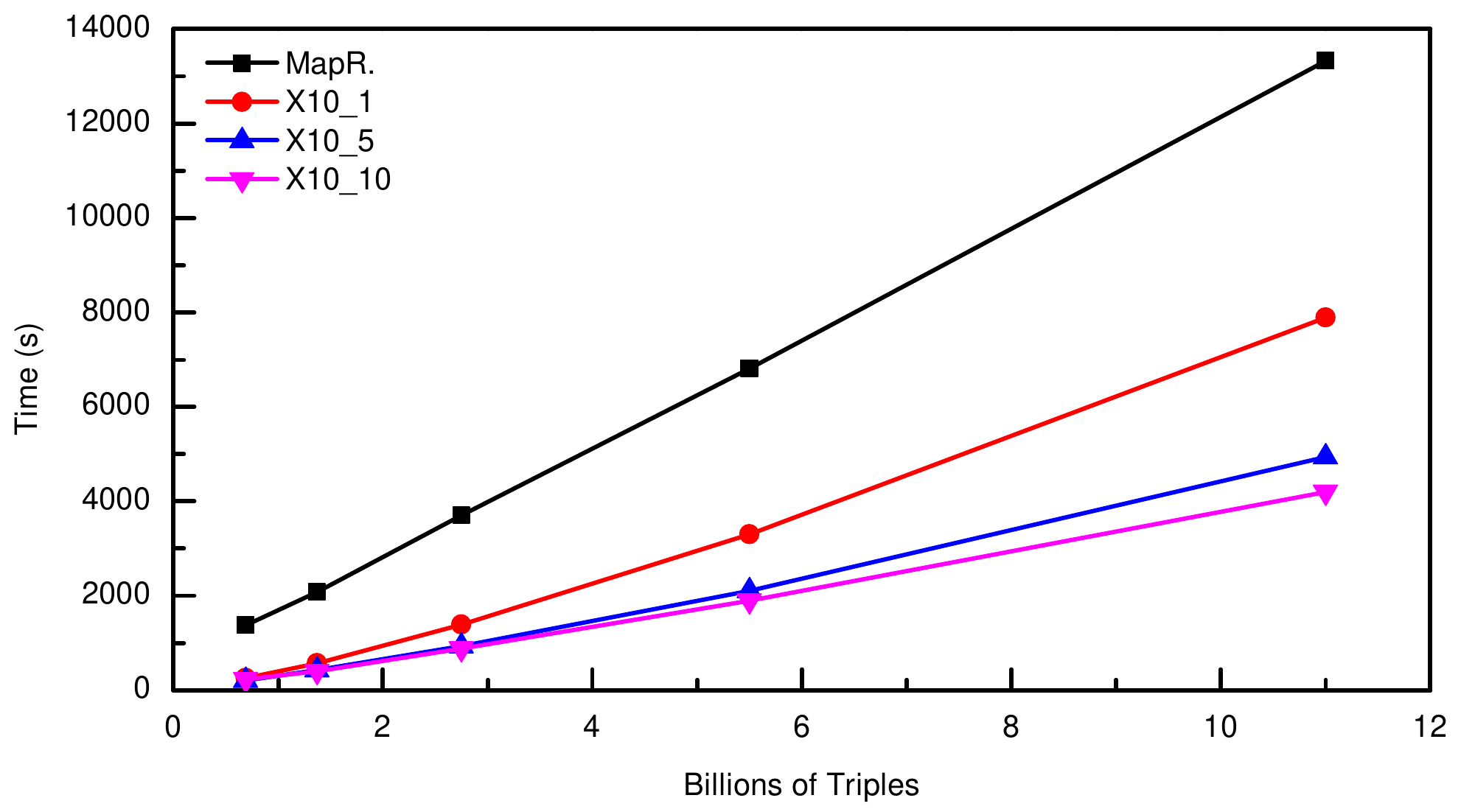} }
\caption{Scalability of two algorithms: (a) encoding 1.1 billion triples with varying the number of computation cores from 12 to 384, (b) the corresponding speedups achieved by varying the cores, and (c) the number of triples starts with 690 million and repeatedly double to 11 billion (192 cores, on disk)}
\label{fig:scalability} 
\end{figure*} 

We test the scalability of our algorithm by varying the number of processing cores and the size of the input data set. We use the LUBM benchmark in our tests as it facilitates the generation of datasets of arbitrary size.

\subsubsection*{Number of Cores} We fix the input data set to 1.1 billion triples and double the number of cores from 12 (single node) till 384. The test results for our algorithm and the MapReduce-based approach are shown in Figure~\ref{fig:subfig:1a}. These results demonstrate that the run time for both algorithms \textit{decreases} with an increase in the number of cores. The speedup obtained with an increasing number of cores compared to a baseline of 12-cores for both algorithms is presented in Figure~\ref{fig:subfig:1b}. In our system, with a small number of cores, the runtime is not linear, since for a single node there is no network communication. Nevertheless, starting from 24 cores, the speedup becomes almost linear (scaled speedup, not shown in the figure, is approximately 1.95). This result supports our theoretical analysis in Section~\ref{sec:algor-compl}, and we attribute the small amount of loss to network traffic. In contrast, the speedup of the MapReduce-based approach is almost linear (even super-linear) initially before plateauing for values of 92 cores and greater. This result mirrors the results obtained in~\cite{Urbani2013}. There can be several reasons for the latter slowdown: we hypothesize that may be due to load imbalance, increased I/O traffic and platform overhead. 

\subsubsection*{Size of Datasets} To study the scalability of our algorithm with increasing input data
size, we create a large LUBM data set with 11 billion triples,
which is roughly equivalent to the \textit{LUBM80000} benchmark. We
split this data set into a number of chunks, each of which contains
140K triples, allowing us to study the effect of \textit{loop} from
Figure~\ref{fig:loop}.

We start our tests with 690 million triples and repeatedly double the
size of the input until we reach a dataset comprising 11 billion triples. Additionally, for each dataset, we also vary the number of \textit{chunks read per loop} for our implementation. The
results are presented in Figure~\ref{fig:subfig:1c}. We see that the runtime for both algorithms is nearly linear with the size of the input data sets. We also notice that MapReduce achieves a slightly super-linear speedup until 5.5 billion triples. After that, MapReduce speedup becomes linear with the input size. For our algorithm, we have experimented with 1, 5, and 10 chunks in each loop.  
One can see that the scalability of our algorithm is
not linear with input data when reading 1 chunk per loop. But,
speedup becomes better as we increase the number of chunks read per loop, and it matches the ideal linear speedup scenario when reading 10
chunks per loop. The reason may be the same as for the transactional case mentioned above, i.e. that a large number for \textit{loop} results in additional runtime overheads as a result of forking threads and object type initializations. Small chunks also results in redundant \textit{filter} and \textit{push} operations for the same terms at the same place in different loops. Such an interpretation is in sympathy with our expectations described in Section~\ref{subsec:flex-memory-footpr}. 

Furthermore, Figure~\ref{fig:subfig:1c} investigates the trade-off between reduced memory consumption and performance as well. For the optimal scalability case with reading 10 chunks at a time, we need to process $10 \times 140K = 1.4M$ triples in each loop. Since, in Table~\ref{tab:1}, we show that 1.1 billion triples is about 190 GB, the size of 1.4 million triples would be about 250 MB, which is well within the RAM availability of most machines. Not withstanding this optimal case implementations using 5 chunks at a time (125 MB) and 1 chunk at a time (25 MB) is only accompanied with little and moderate scalability loss respectively.

\subsection{Load Balancing}
\label{sec:load-balancing}

We measure the load-balance characteristics of our algorithm in terms of five metrics defined later in this section. We instrument our code with counters to gather data for the first four metrics. The data for the final metric is obtained using the tracing option provided by the X10 implementation.

\begin{itemize}
\item \emph{number of outgoing terms}: The number of terms transferred to a remote place. This metric gives insight into the \textit{communication}
load balance achieved by our algorithm. For example, the larger the number of outgoing terms, the greater the associated network traffic.
\item \emph{number of misses}: The number of terms that are not already encoded (\textit{missed}) in the dictionary and hence require the generation of a new id.
\item \emph{miss ratio}: The number of misses divided by the sum of hit and miss for the local dictionary.
\item \emph{number of processed terms}: the number of terms processed
  by a computing node.
\item \emph{received bytes}: the size of processed terms in bytes at a
  computing node.
\end{itemize}

We encoded 1.1 billion LUBM triples on a varying number
of cores to gather data for the first three metrics described above. The results are presented in Table~\ref{tab:5}. We can see that the average values of the three metrics for all the tests are very close to the maximum values, suggesting excellent load balancing performance. The scalability of our algorithm with an increasing number of processing cores is highlighted well in these results. There is a clear linear decrease in all three metrics with an increase in the number of processing cores. Finally, the results also illustrate a consistent almost uniform miss probability for each dictionary.  The average miss ratio is about 94.5\%, indicating that we have redundant computation on average for 5 out of every 100 terms. This ratio approached the ideal value of 100\%, which is nevertheless difficult to achieve in a distributed systems without significant coordination overhead. Additionally, our implementation is still based on the \textit{all-to-all} communication, which could possibly effect the performance. However, our system does not repartition all the data, but only transfers the mappings that are necessary for each node. In this sense, our system performs useful computation in terms of data locality in 94.5\% of the cases, meaning that although our approach does require communication between all nodes, only moving the data that actually needed.

\begin{table}[!t]
  \centering
  \caption{Term Information during encoding 1.1 billion triples}
  \begin{tabular*}{\columnwidth}{@{\extracolsep{\fill}}ccccccc}
    \toprule[0.8pt]
    \multirow{2}[4]{*}{\textbf{\# Core}} & \multicolumn{2}{c}{\textbf{\# Outgoing (M)}} & \multicolumn{2}{c}{\textbf{\# Misses (M)}} & \multicolumn{2}{c}{\textbf{Miss Ratio}} \\
    & \textbf{Max} & \textbf{Avg.} & \textbf{Max} & \textbf{Avg.} & \textbf{Max} & \textbf{Avg.} \\
    \midrule
    24    & 11.65 & 11.59 & 10.95 & 10.95 & 95.7\% & 94.5\% \\
    48    & 5.85  & 5.78  & 5.46  & 5.46  & 96.1\% & 94.5\% \\
    96    & 2.94  & 2.89  & 2.73  & 2.73  & 96.1\% & 94.5\% \\
    192   & 1.48  & 1.43  & 1.35  & 1.35  & 96.4\% & 94.5\% \\
    384   & 0.74  & 0.70  & 0.90  & 0.87  & 96.4\% & 94.5\% \\
    \bottomrule[0.8pt]
  \end{tabular*}%
  \label{tab:5}%
\end{table}%

\begin{table}[!t]
  \centering
   \caption{Comparison of Received Data for each computing node when processing 1.1 billion triples using 192 cores (In millions)}
  \begin{tabular*}{\columnwidth}{@{\extracolsep{\fill}}cccccc}
    \toprule[0.8pt]
    \multicolumn{2}{c}{\multirow{2}[4]{*}{\textbf{Algorithm}}} & \multicolumn{2}{c}{\textbf{Recv. Bytes}} & \multicolumn{2}{c}{\textbf{Recv. Records}} \\
    \multicolumn{2}{c}{} & \textbf{Max.} & \textbf{Avg.} & \textbf{Max.} & \textbf{Avg.} \\
        \midrule
    \multirow{3}[6]{*}{\textbf{MapR.}} & \textbf{Job1} & 9.94  & 4.02  & 24.04 & 1.73 \\
          & \textbf{Job2} & 135.61 & 79.77 & 30.91 & 17.28 \\
          & \textbf{Job3} & 120.81 & 106.82 & 19.61 & 17.28 \\
              \midrule
    \multicolumn{2}{c}{\textbf{X10}} & 194.71 & 187.82 & 1.48  & 1.43 \\
    \bottomrule[0.8pt]
  \end{tabular*}%
  \label{tab:6}%
\end{table}%

The last two metrics capture the load at each compute node in terms of the number of terms processed and size of data received in bytes. These metrics are important for measuring computational load balance and are used here to provide comparison with the performance available using the MapReduce approach. Since MapReduce divides the whole compression into three separate jobs and the implementation does not provide the relative metrics, we extract the \emph{reduce input records} and
\emph{reduce shuffle bytes} in the reduce phase of each job from the
Hadoop logs. These two items indicate the number of records processed and the corresponding data sizes for each of the 192 reduce tasks.

The results are summarized in Table~\ref{tab:6} and demonstrate that the difference between the maximum and the average value of these metrics for our implementation is much smaller than MapReduce, indicating better load balancing (in addition to the results, the minimum number of bytes received is $184.70M$ and the minimum number of records received is $1.37M$ in our approach, also showing minimal skew). Furthermore, when comparing the sum total of bytes received across the two implementations, it is clear that our proposed technique results in better performance. Consequently even when comparing with the reduce phase of MapReduce, our system results in a lighter workload and less network communication.

\section{Conclusions and Future Work}
\label{sec:concl-future-work}

In this paper, we have introduced a dictionary encoding algorithm for
the compression of big RDF data. The algorithm utilises the X10 system
which is based on the APGAS programming model. Using the X10 language,
and in turn the APGAS model, has a number of advantages: (a) flexible
and efficient scheduling. APGAS, like PGAS, separates tasks from
underlying concurrency model, thereby allowing one to implement an
efficient scheduling strategy irrespective of the number of tasks forked
using \texttt{async}. (b) APGAS being derived from both MPI and OpenMP
programming models, extracts parallelism at both the distributed and
single machine hierarchies. (c) Finally, an abstract model provided by
the \texttt{async}, \texttt{finish}, \texttt{places}, and
\textit{activities}, helps one write short code, which is easier to
debug and maintain.

We have presented an extensive quantitative evaluation of the proposed
algorithm and conducted a comparison with a state-of-art system using
the MapReduce model. Our main conclusions are that the proposed
algorithm is: (a) Highly scalable both with increments in number of
cores and in the size of the dataset, (b) Computationally fast,
encoding 11 billion statements in about 1.2 hours, and achieving a
$2.6 -7.4\times$ improvement over the MapReduce method, (c) Flexible for
various semantic application scenarios, (d) Robust against data skew,
showing excellent load balancing, and (e) Suitable for use and further
development as part of a high performance distributed system.


X10 can be compiled and run on GPUs \cite{Cunningham2011}. Future work
will focus on the exploitation of this technology to achieve even higher
performance. Our long term goal is to develop a highly scalable data
distribution management system for extreme scale RDF data.



\bibliographystyle{IEEEtran}
\bibliography{IEEEabrv,IPDPS}

\end{document}